# Student–ChatGPT Interaction Visible: Designing a Teacher Dashboard for EFL Writing Education

**Minsun Kim, Seon Gyeom Kim, Suyoun Lee, Yoosang Yoon, Junho Myung, Haneul Yoo, Jieun Han, Hyunseung Lim, Yoonsu Kim, So-Yeon Ahn, Juho Kim, Alice Oh, Hwajung Hong, Tak Yeon Lee**
KAIST, Korea Advanced Institute of Science and Technology[1]
{9909cindy, takyeonlee}@kaist.ac.kr

**ABSTRACT**: We present a Prompt Analytics Dashboard (PAD) for teachers that can traces student-LLM interactions from EFL writing classes. PAD can show student prompt-response exchanges with LLM chatbot and English essay writing revision histories to support data-informed instruction and visibility in classes. Through two iterative co-design sessions with six EFL instructors, we distilled a compact trace taxonomy (misuse signals, goal-alignment cues, revision effort) and instantiated three interface views (overview, week/outcome filter, drill-down with evidence snippets). This pipeline summarizes potential misuse and alignment at class/cohort levels and attaches micro-explanations to reduce over-surveillance. Instructors reported reduced scanning burden and clearer timing for interventions.

**Keywords**: learning analytics, teacher dashboard, L2 writing, large language models, prompt analysis, misuse detection

## 1 BACKGROUND AND GAP

Recent advancements in Large Language Models (LLMs) like ChatGPT have transformed education by enabling personalized learning. In EFL contexts, ChatGPT provides instant feedback, examples, and writing support (Sabzalieva & Valentini, 2023), though concerns remain about overreliance, plagiarism, and ethical risks (Rahman & Watanobe, 2023). However, beyond the availability of tools, instructors face a pedagogical challenge: student–LLM interactions generate large volumes of data that are difficult to interpret within limited instructional time. Without analytic support, teachers struggle to identify when and how to intervene meaningfully in students' AI-supported writing processes. PAD is designed to address this challenge by transforming raw interaction traces into interpretable signals that support teachers' instructional decision-making, such as identifying overreliance of LLM support, misalignment with learning objectives, or productive revision effort.

## 2 TWO-PHASE ITERATIVE CO-DESIGN PROCESS

### 2.1 Phase 1: Understanding Teachers' Needs and Practices

In the first phase, we conducted surveys and interviews with six university EFL instructors who integrated ChatGPT into their writing courses. The goal was to identify teachers' perceptions, challenges, and design requirements for PAD. Teachers evaluated ChatGPT's educational usefulness and potential risks (e.g., plagiarism, bias, overreliance). We also observed how teachers reviewed students' essay drafts and ChatGPT chat logs to assess learning effectiveness. Insights from this phase revealed design goals: (1) Providing a quick overview of student-ChatGPT interactions, (2) Identifying common patterns and undesirable usage of ChatGPT, (3) Facilitating in-depth analysis of essay editing and chat log, and (4) Customizing prompt instructions for personalized learning environment.

[1] This work was mainly supported by Elice, a leading company in the domain of digital education. The Azure credits for hosting the ChatGPT service were supported by Microsoft Accelerate Foundation Models Research (AFMR) grant program.

### 2.2 Phase 2: Prototype Co-Design and Evaluation

Based on Phase 1 findings, we developed a PAD prototype visualizing students' essay, ChatGPT interactions, and automated analytics (e.g., misuse-related analytic cues, learning-objective alignment). In Phase 2, teachers interacted with the prototype and participated in semi-structured interviews to provide feedback on usability, interpretability, and pedagogical potential. Thematic analysis of interview data revealed four key design implications: (1) simplifying analytic interpretations on general statistics, (2) improving teachers' efficiency to browse log, (3) enhancing instructors' ability to provide personalized, targeted feedback, and (4) supporting flexible, teacher-driven customization on dashboard design.

## 3 PROMPT ANALYTICS DASHBOARD

### 3.1 System Architecture

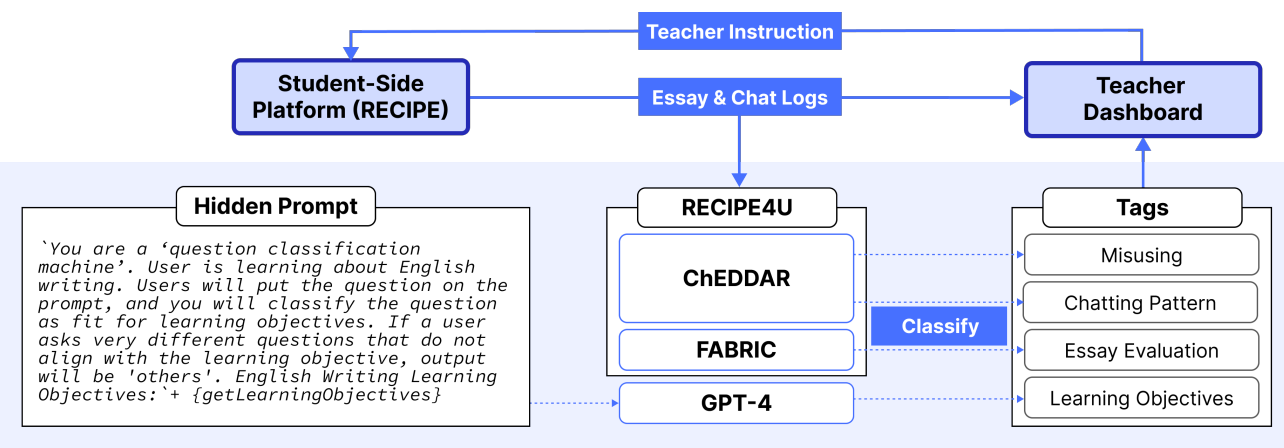

**Figure 1: Screenshot of overall system architecture.**

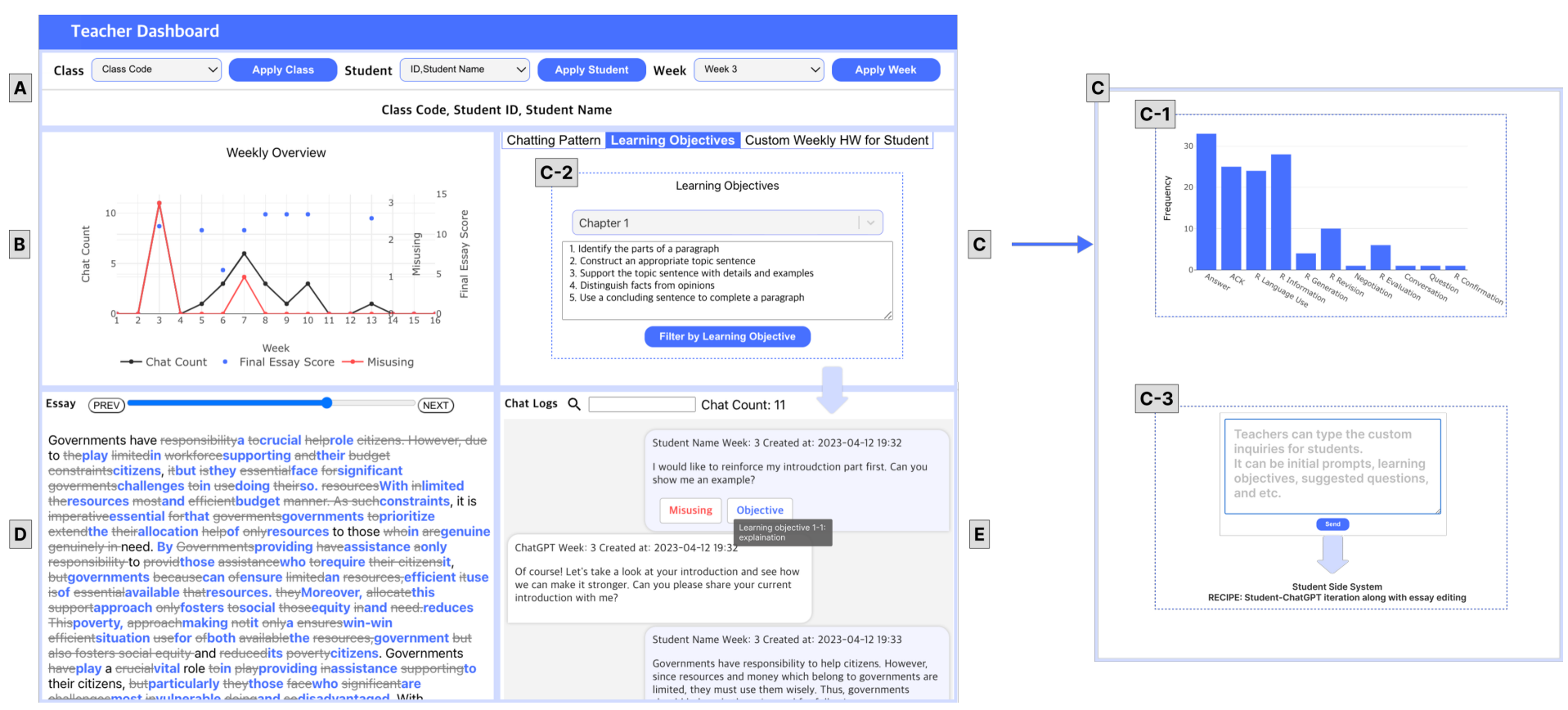

**Figure 2: Screenshot of dashboard prototype.**

The dashboard is integrated within a larger system where students write weekly essays and interact with ChatGPT through a student-side application. Figure 2 shows three main components of PAD. The student-side platform (Han & Yoo, 2023) provides a web-based essay writing interface where students compose and revise their essays with ChatGPT API support. Their essays and chat logs are analyzed by three LLM-based components: (1) GPT-4 surface potential misuse [2] signals (e.g., requests for generation without revision and chat patterns (e.g, negotiations, asking for essay evaluation), (2) automated essay scoring model(Han & Yoo, 2023), and (3) GPT-4 classifies prompts by learning objectives.

### 3.2 Feature Description

The design follows "overview–zoom–filter–details-on-demand" to balance macro-level monitoring and micro-level contextual analysis (Shneiderman, B., 2003). Figure 2 shows the screenshot of dashboard prototype. (A) Population pool: The dashboard displays information calculated based on the selected population pool. Teachers can narrow the analysis pool by selecting classes, students, and weeks. (B) Overview: This is an overview chart with the x-axis fixed to weekly, allowing you to see the chat count over time, the last saved essay score graded by AI, and the misuse count. (C) This area consists of three tabs, (C-1) a chart for chatting patterns, (C-2) additional filtering options based on learning objectives, and (C-3) the ability to deliver messages to students. (D) Teachers can track changes to essays by moving sliders. (E) Shows the students' entire chat history. Teachers can search, and each prompt has a tag so teachers can check the chat content.

## 4 CONCLUSIONS AND IMPLICATIONS

This study introduced the Prompt Analytics Dashboard (PAD), a teacher-centered tool designed to visualize and interpret students' ChatGPT-assisted writing activities. Through an iterative co-design process, we identified teachers' needs for efficiency, pedagogical visibility, and ethical AI integration. PAD demonstrates how learning analytics can extend beyond performance tracking to support interpretive, pedagogical decision-making in AI-mediated classrooms. The findings highlight the importance of human-centered dashboard design that empowers teachers to guide responsible LLM use. Future work should explore scalability, real-time feedback, and adaptive analytics to further enhance data-informed teaching practices.

[2] In PAD, misuse does not denote definitive judgments of academic misconduct; rather, it refers to teacher-defined patterns of AI use that may pose pedagogical risks and therefore warrant closer instructional attention.